\documentclass[%
 reprint,
 showpacs,
 amsmath,amssymb,
]{revtex4-1}

\usepackage{dcolumn}
\usepackage[dvipdfmx]{graphicx}
\usepackage{mathrsfs}
\usepackage{accents}
\usepackage{bm}
\usepackage{color}
\usepackage{here}

\makeatletter
\def\btt#1{\texttt{\@backslashchar#1}}
\DeclareRobustCommand\bblash{\btt{\@backslashchar}} \makeatother

\begin{document}

\title{Tunable $\varphi$-Josephson junction with a quantum anomalous Hall insulator }

\author{Keimei Sakurai$^1$, 
Satoshi Ikegaya$^1$,  
and Yasuhiro Asano$^{1,2,3}$}
\affiliation{$^1$Department of Applied Physics,
Hokkaido University, Sapporo 060-8628, Japan}
\affiliation{$^2$Center for Topological Science and Technology,
Hokkaido University, Sapporo 060-8628, Japan}
\affiliation{$^3$Moscow Institute of Physics and Technology, 
141700 Dolgoprudny, Russia}
\date{\today}

\begin{abstract}
 We theoretically study the Josephson current in a superconductor/
quantum anomalous Hall insulator/superconductor junction 
 by using the lattice Green function technique. 
When an in-plane external Zeeman field is applied to the quantum 
anomalous Hall insulator, 
the Josephson current $J$ flows without a phase difference across the junction $\theta$. 
The phase shift $\varphi$ appearing in the current-phase relationship $J \propto \sin(\theta-\varphi$) 
 is proportional to the amplitude of Zeeman fields and depends on the direction of Zeeman fields. 
A phenomenological analysis of the Andreev reflection processes explains the physical origin of $\varphi$. 
In a quantum anomalous Hall insulator, time-reversal symmetry and mirror reflection symmetry are broken
simultaneously. However magnetic mirror reflection symmetry is preserved. Such characteristic symmetry property enable us to have a tunable $\varphi$-junction with a quantum Hall insulator. 
\end{abstract}


\maketitle

\section{Introduction}
The Josephson effect is a macroscopic quantum phenomenon caused by the spatial 
fluctuations of superconducting phase~\cite{josephson}. 
 When two superconductors (S) sandwich a material X, the Josephson current $J$ flows 
as a function of the phase difference across the junction $\theta$. 
 The current-phase $(J-\theta)$ relationship (CPR) reflects well the electronic properties of 
 X~\cite{likharev,A.A.Golubov}.
When X is an insulator, the CPR is sinusoidal 
$J =J_0 \sin\theta$ with $J_0>0$ being the critical curent~\cite{ambegaokar}. 
Such junction is called 0-junction because the junction energy is minimum at the zero phase difference.
A $\pi$-junction in which the energy is minimum at $\theta=\pm \pi$ can be realized when 
we choose a ferromagnet as X~\cite{buzdin,buzdin2}.
The spin-singlet pairing correlation spatially oscillates and changes its sign 
under the exchange potential. Therefore such Josephson junction undergoes the transition 
between the 0-state and the $\pi$-state alternatively as the variation of the thickness in the ferromagnet~\cite{ryazanov,kontos}.
In the view of the device application, a $\pi$-junction plays a crucial role in constructing a flux qubit.
Choosing an insulating ferromagnet as X makes a devise with a long coherence time possible~\cite{SFIS}.

The energy of a Josephson junction some of the time takes
its minimum at a phase difference $\varphi$ which is neither 0 nor $\pi$.
The CPR in such $\varphi$-junction $J=J_0 \sin(\theta-\varphi)$ suggests that 
the current flows even at the zero phase difference~\cite{Jun-Feng_PRB,T.Yokoyama_PRB_2014,A.Rasmussen_PRB}.
Breaking time-reversal symmetry in X is a necessary condition to realize the $\varphi$-junction.
The value of $\varphi$ is determined by characteristic electronic structures in X. 
So far, the possibility of $\varphi$-junction has been discussed theoretically in various Josephson junctions with 
X being multilayered ferromagnets~\cite{SFFFS,SNFS}, quantum point contacts~\cite{QPC}, quantum dots~\cite{QD1,QD2,QD3}, nanowires~\cite{T.Yokoyama_PRB_2014,NW2}, topological materials~\cite{STIS,SQSHIS}, and a ferromagnet without inversion 
symmetry~\cite{buzdin3}. 
In experiments, on the other hand, the realization of a $\varphi$-junction has been reported only in a 
Josephson junction with a nanowire quantum dot~\cite{EOF}.
At present, it is not easy to controle the phase shift $\varphi$ after fabricating Josephson junctions. 

In this paper, we study the Josephson effect in superconductor/
quantum anomalous Hall insulator/superconductor (S/QAHI/S) 
junction theoretically. 
A QAHI is a topologically nontrivial material in two-dimension and breaks 
time-reversal symmetry by its spontaneous magnetization. 
In experiments, doping of magnetic elements such as
Cr~\cite{CrTI1,CrTI2,CrTI3,CrTI4} and V~\cite{VTI} onto a thin film of a topological insulator 
$(\mathrm{Bi},\mathrm{Sb})_2 \mathrm{Te}_3$ enables QAHIs.
According to the bulk-boundary correspondence, 
nonzero Chern number implies the presence of chiral edge states. 
We will discuss characteristic features in the Josephson current flowing through such a chiral edge channel. 
The Josephson current is calculated numerically by using the lattice Green function method. 
When we apply an in-plane external Zeeman field to QAHI, the junction becomes a $\varphi$-junction.
Moreover, the value of $\varphi$ is proportional to a Zeeman field, which suggests a possibility
of tunable $\varphi$-junction. 
A phenomenological argument explains well the physical origin of the $\varphi$-junction.
The breaking magnetic mirror reflection symmetry of the Hamiltonian is a key 
property to understand the physics behind the phase shift $\varphi$. 
We also demonstrate that random impurities and the asymmetry of junction geometry in real space 
 break magnetic mirror reflection symmetry and make S/QAHI/S be a $\varphi$-junction.

This paper is organized as follows. In Sec.~II, we show the Hamiltonian of a QAHI 
on tight-bainding model and discuss the numerical results of the Josephson current.
In Sec.~III, the mechanism of the $\varphi$ phase shift in the CPR is explained 
by a phenomenological analysis of the Andreev reflection processes. 
The numerical results in the presence of random impurities and junction asymmetry 
are presented in Sec.~IV. The conclusion is given in Sec.~V.

\section{Numerical results on a tight-binding model}

\begin{figure}[t]
  \begin{center}
    \includegraphics[clip,scale=0.40]{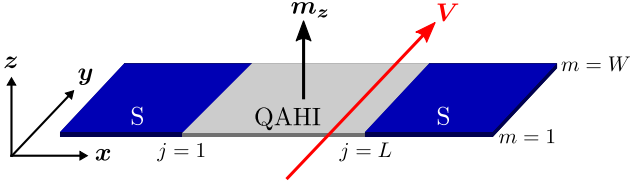}
  \caption{Schematic figure of a Josephson junction with a QAHI. 
  The magnetic moment $\boldsymbol{m_{\mathrm{z}}}$ points in the $z$ direction.
An external Zeeman field $\boldsymbol{V}$ is applied within the two-dimensional plane.}
      \label{fig:model}
  \end{center}
\end{figure}

Let us consider a superconductor/quantum anomalous Hall insulator/superconductor (S/QAHI/S) junction 
on a two-dimensional tight-binding model as shown in Fig.~\ref{fig:model}. 
Throughout this paper, we measure the length in unit of the lattice constant. 
A vector $\bm{r} = j\bm{x} + m\bm{y}$ points a lattice site, where $\bm{x}$ and $\bm{y}$ are 
the unit vectors in the $x$ and $y$ directions, respectively.
The junction consists of three regions: a quantum anomalous hall insulator( i.e., $1 \leq j\leq L$ ) and two superconductors ( i.e., $-\infty \leq j \leq 0$ and $L+1 \leq j \leq \infty$ ). 
An external Zeeman field $\boldsymbol{V}$ is applied in the QAHI segment. 
The width of the junction is $W$. We apply the hard wall boundary condition in the $y$ direction. 
The Hamiltonian of the junction is given by
 \begin{align}
  \mathcal{H}
   =& 
    \mathcal{H}_{\mathrm{L}} +
    \mathcal{H}_{\mathrm{QAHI}} +
    \mathcal{H}_{\mathrm{R}}.
\label{TH}
 \end{align}
The first and the third term in Eq.~(\ref{TH}) are the Hamiltonians of a $s$-wave superconductor 
on the left and that on the right, respectively. They are given by
\begin{align}
  \mathcal{H}_{\mathrm{L}(\mathrm{R})}
    =&
    \sum_{\boldsymbol{r},\boldsymbol{r'}}
    \Psi^{\dagger}_{\boldsymbol{r}}
    \left[
    \begin{array}{cc}	
    \hat{h}_{\boldsymbol{r},\boldsymbol{r'}}  &
    {\hat\Delta}_{\boldsymbol{r},\boldsymbol{r'}}
    \mathrm{e}^{i\theta_{L(R)}} \\
    -\hat{\Delta}_{\boldsymbol{r},\boldsymbol{r'}}
    \mathrm{e}^{-i\theta_{L(R)}} &
    -\hat{h}^*_{\boldsymbol{r},\boldsymbol{r'}}  
    \end{array}
    \right]
    \Psi_{\boldsymbol{r'}}, \\
    %
    %
  \hat{h}_{{\boldsymbol{r},\boldsymbol{r'}}}
    =& \left[  
    -t\delta_{\vert \boldsymbol{r}-\boldsymbol{r'} \vert,1}
    + (4t - \mu_{\mathrm{s}}) \delta_{\boldsymbol{r},\boldsymbol{r'}}
    \right] \hat{\sigma}_0, \\
  \hat{\Delta}_{\boldsymbol{r},\boldsymbol{r'}}  
    =&i\, \Delta \,
      \hat{\sigma}_2 \, \delta_{\boldsymbol{r},\boldsymbol{r'}},\\
   \Psi_{\boldsymbol{r}} = &\left(
      c_{\boldsymbol{r},\uparrow}, \,
      c_{\boldsymbol{r},\downarrow}, \, 
      c^{\dagger}_{\boldsymbol{r},\uparrow}, \,
      c^{\dagger}_{\boldsymbol{r},\downarrow} 
   \right)^\mathrm{T},
  \end{align}
where $c^{\dagger}_{\boldsymbol{r},\sigma}$ ($c_{\boldsymbol{r},\sigma}$) is the creation (annihilation) operator of an 
electron at $\boldsymbol{r}$ with spin $\sigma$($=\uparrow$ or $\downarrow$), 
$\hat{\sigma}_j$ with $j=1$-$3$ are the Pauli matrices in spin space,  
$\hat{\sigma}_0$ is the unit matrix,  
$\mu_{\mathrm{s}}$ is the chemical potential in the superconductors, $\mathrm{T}$ means the transpose of a matrix,
and $\Delta$ is the amplitude of the pair potential. We consider the hopping integral $t$ between the nearest-neighbor sites. 
The phase of the left (right) superconductor is $\theta_L$ ($\theta_R$). The physical values depends only on the 
phase difference across the junction $\theta=\theta_L-\theta_R$.

The second term in Eq.~(\ref{TH}) indicates the Hamiltonian of a QAHI~\cite{Model}
\begin{align}
 \mathcal{H}_{\mathrm{QAHI}}
   =&
   \sum_{\boldsymbol{r},\boldsymbol{r'}}
   \Psi^{\dagger}_{\boldsymbol{r}}
   \left[
   \begin{array}{cc}
     \hat{Q}_{\boldsymbol{r},\boldsymbol{r'}} &  0
   \\  0                  &-\hat{Q}^*_{\boldsymbol{r},\boldsymbol{r'}}   
   \end{array}
   \right]
   \Psi_{\boldsymbol{r'}},\label{HQ} \\
 \hat{Q}_{\boldsymbol{r},\boldsymbol{r'}}
   =& \left[ 
   -t \delta_{\vert \boldsymbol{r}-\boldsymbol{r'} \vert,1}
   + (4t-m_{\mathrm{z}})\delta_{\boldsymbol{r},\boldsymbol{r'}}
   \right] \hat{\sigma}_{3}
   \nonumber\\ &
    -\dfrac{i\lambda}{2}
    \left[
    \delta_{j,j'+1}-\delta_{j+1,J'}
    \right] \delta_{m,m'} \hat{\sigma}_{2}
    \nonumber\\ &
	+\dfrac{i\lambda}{2}
	\left[
    	\delta_{m,m'+1}-\delta_{m+1,m'}
	\right] \delta_{j,j'} \hat{\sigma}_{1}
	\nonumber\\ &
	-V_x \delta_{\boldsymbol{r},\boldsymbol{r'}} \hat{\sigma}_{1}
	-V_y \delta_{\boldsymbol{r},\boldsymbol{r'}} \hat{\sigma}_{2}
    ,\label{QR}
 \end{align}
where $\lambda$ is the amplitude of the spin-orbit interaction and $m_{\mathrm{z}}$ is a Zeeman potential 
induced by the spontaneous magnetization. 
For $m_{\mathrm{z}} > 0$, a QAHI has a chiral edge state characterized by a Chern number of $\mathbb{Z}=-1$~\cite{TKNN}.
The insulating gap can be described by these parameters as 
$E_{\mathrm{g}} \sim 2\lambda \sqrt{m_\mathrm{z}/{t}}$. (See also Appendix A for details.) 
In addition to the spontanesous magnetic moment in the $z$ direction, 
we consider Zeeman potentials $V_x$ in the $x$ direction and $V_y$ in the $y$ direction by
applying an external magnetic field. 
In what follows, we assume weak Zeeman field so that
$ \vert V_x \vert\ll m_{\mathrm{z}}$ and $\vert V_y \vert \ll m_{\mathrm{z}}$ are satisfied. 

We calculate the Josephson current based on the current formula~\cite{A.Furusaki,Y.Asano_PRB_2001,Y.Asano}
 \begin{align}
   J =& \dfrac{ie}{2\hbar}T \sum_{\omega_n} \mathrm{Tr}
     \left[
     \check{\tau}_3 \check{T}_{+}
     \check{G}(\boldsymbol{r},\boldsymbol{r}+\boldsymbol{x};\omega_n) \right.\nonumber\\
&\left.      - \check{\tau}_3 \check{T}_{-}
     \check{G}(\boldsymbol{r}+\boldsymbol{x},\boldsymbol{r};\omega_n)   
     \right],\label{JC}
 \end{align}
 \begin{align}
   \check{T}_{\pm}
     =&
    \left[
    \begin{array}{cc}
    \hat{t}_{\pm} &
    0            \\
    0             &
    -\hat{t}^*_{\pm}
    \end{array}
    \right], \quad
  \hat{t}_{\pm}
    =&
    \left[
    \begin{array}{cc}
    - t \bar{1}                     &
    \mp \dfrac{\lambda}{2} \bar{1}  \\
    \pm \dfrac{\lambda}{2} \bar{1}  &
      t \bar{1}                     
    \end{array}
    \right],
 \end{align}
where $\check{G}(\boldsymbol{r},\boldsymbol{r}';\omega_n)$ is the Matsubara Green function and 
$\omega_n=(2n+1)\pi k_B T$ is the Matsubara frequency with $n$, $T$ and $k_B$ being an integer number, a temperature 
and the Boltzmann constant, respectively. 
The Green function is calculated numerically by using the lattice Green function technique~\cite{lee}.
In above equations, $\hat{\cdots}$($\check{\cdots}$) indicates $2W \times 2W$ ($4W \times 4W$) matrices, 
and the $W \times W$ unit matrix is denoted by $\bar{1}$.
In Eq.~(\ref{JC}), $\check{\tau}_3 $ is the third Pauli matrix in particle-hole space, and  
$\mathrm{Tr}$ means the trace over spin space, particle-hole space, and the summation 
over the lattice sites in the $y$ direction.

%
%
%
\begin{figure*}[htbp]
  \begin{center}
   \includegraphics[clip, scale=0.52]{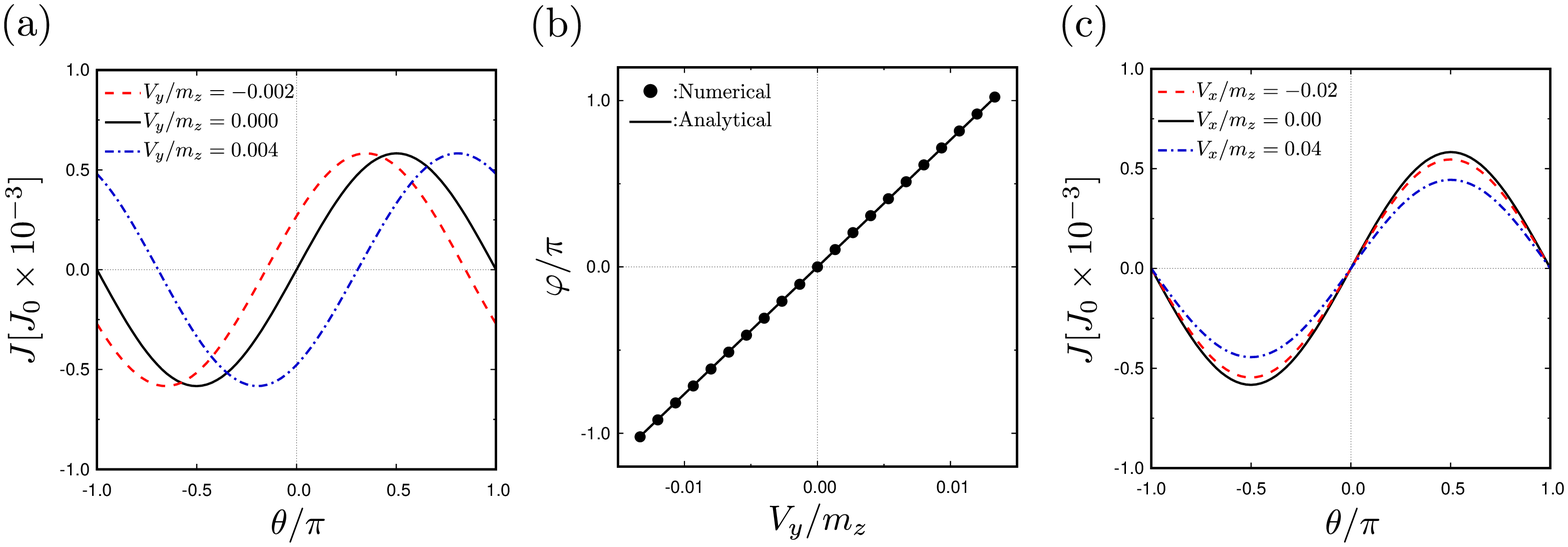}
\caption{(a) Josephson current as a function of $\theta$ at $V_x = 0$. 
(b) The relation between the phase shift $\varphi$ and $V_y$. The results obtained in numerical simulation are shown with 
symbols. A solid line indicates the analytical results in Eq.~(\ref{shift_v}), where we use $\lambda=t$ and $L=80$ 
consistently with the numerical simulation.
(c) Josephson current as a function of $\theta$ at $V_y = 0$.    
}
    \label{fig:main}
 \end{center}
\end{figure*}

Throughout this paper, we fix basic parameters as $m_z=1.5t$ and $\lambda = 1.0t$. 
The chiral edge states spatially localize within 2 lattice constants from the surface under these parameter choice. 
The size of a QAHI should be much larger than the localization length.
Therefore, we choose $L=80$ and $W=20$.
The chemical potential in two superconductors is chosen as $\mu_s=2.7 t$. 
We find that $\mu_s >m_z$ is necessary so that the normal conductance through the chiral edge is quantized 
at $e^2/h$. Otherwise, the normal conductance becomes smaller than $e^2/h$.
The amplitude of the pair potential at the zero temperature $\Delta_0=0.001t$ is much smaller than $\mu_s$. 
We mainly calculate the Josephson current at $T=0.1T_c$, where $T_c $ is 
the transition temperature.
We assume that the superconducting gap is smaller than the insulating gap in a QAHI (i.e., $\Delta_0 \ll E_{\mathrm{g}}$).
As a result, the Josephson current flows only through the 
chiral edge states in a QAHI.

In Fig.~\ref{fig:main} (a), we first discuss the CPR in the absence of an external Zeeman field as shown with a solid line.
The Josephson currents are normalized to $J_0 \times 10^{-3}$ with $J_0 = e\Delta_0/\hbar$. 
The CPR at $V_y=0$ in Fig.~\ref{fig:main} (a) is sinusoidal, which is a robust feature independent of the parameters such as 
$T$, $\mu_s$, and $m_z$.
In Appendix B, we demonstrate the CPR for several choices of $\mu_s$ to check the robustness of 0-junction 
in the absence of the Zeeman field. 
Although the dependence of the Josephson critical 
current on temperatures indicates unusual behavior as shown in Appendix B, we confirmed that 
the CPR is always sinusoidal in the absence of Zeeman fields.

When we apply a Zeeman field in the $y$ direction, the results for $V_y/m_z=0.004$ and $-0.002$ in Fig.~\ref{fig:main}(a) 
deviate from the sinusoidal relation. A $\varphi$-junction is realized by applying a Zeeman field in the $y$ direction.
We also find that the amplitude of the critical current is insensitive to $V_y$. The phase shift $\varphi$ 
in the CPR $J=J_c\sin(\theta-\varphi)$ is plotted as a function of $V_y$ in Fig.~\ref{fig:main}(b) with symbols. 
The results suggest a linear relationship between $\varphi$ and $V_y$. Namely the phase shift $\varphi$ is 
tunable in a S/QAHI/S junction.

 Figure~\ref{fig:main}(c) shows the numerical results of the Josephson current 
 under Zeeman fields in the $x$ direction.
For both $V_x/m_z=0.04$ and $-0.02$, the CPR is sinusoidal. 
Thus the phase shift depends on the direction of a Zeeman field.
These findings are the central results of this paper.
In the next section, 
we will explain the mechanism for the phase shift and its anisotropic response to the Zeeman field 
by considering the Andreev reflection processes through the chiral edge states in a QAHI.

\section{Origin of phase shift}
\subsection{Symmetry analysis}
To discuss the mechanism for the phase shift $\varphi$ in the CPR, we first analyze the symmetry of 
Hamiltonian.
For this purpose, we describe the Hamiltonian in continuas space as 
\begin{align}
&H_0(\theta) = 
    H_{\mathrm{L}} + H_{\mathrm{QAHI}}+ H_{\mathrm{R}},\\
  &H_{\mathrm{L}(\mathrm{R})}(\boldsymbol{r})
    =    
    \left[
    \begin{array}{cc}
    \hat{h}(\boldsymbol{r})                      &
    \Delta_0 i \hat{\sigma}_2 \mathrm{e}^{i\theta_{L(R)}} \\
    -\Delta_0  i \hat{\sigma}_2 \mathrm{e}^{-i\theta_{L(R)}}  &
    -\hat{h}^* (\boldsymbol{r}) 
    \end{array}
    \right], \label{LS} \\
  &H_{\mathrm{QAHI}}(\boldsymbol{r})
    = 
    \left[
    \begin{array}{cc}
    \hat{Q}(\boldsymbol{r})         &
    0                \\
    0                &
    -\hat{Q}^*(\boldsymbol{r})  
    \end{array}
    \right], \label{QAHI}
\end{align}
with 
\begin{align}
  \hat{h}(\boldsymbol{r}) =& ( \varepsilon_{\boldsymbol{r}} - \mu_s ) \hat{\sigma}_0, \quad
  \varepsilon_{\boldsymbol{r}}= -\dfrac{\hslash^2 }{2m}
       \nabla^2,\\
  \hat{Q} (\boldsymbol{r})
    =& (\varepsilon_{\boldsymbol{r}} - m_z)\hat{\sigma}_3
    +i\lambda \partial_x \hat{\sigma}_2
    -i\lambda \partial_y \hat{\sigma}_1. \label{Q}	   
\end{align}
The Hamiltonian for a Zeeman field is given by
\begin{align}
 H_{\mathrm{Zeeman}}
   =&  
    H_{V_x} + H_{V_y},\\
H_{V_x}=&  
    \left[
    \begin{array}{cc}
    -V_x \hat{\sigma}_1         &
    0                \\
    0                &
    V_x \hat{\sigma}_1  
    \end{array}
    \right], \\
H_{V_y}=&  
    \left[
    \begin{array}{cc}
    - V_y \hat{\sigma}_2         &
    0                \\
    0                &
    - V_y \hat{\sigma}_2  
    \end{array}
    \right].
\end{align}

In the Hamiltonian of a QAHI in Eq.~(\ref{Q}), both mirror reflection symmetry with respect to the 
$xz$-plane and time-reversal symmetry are broken simultaneosly. 
These facts are represented by the relations
\begin{align}
\mathcal{M}_{xz} H_{\mathrm{QAHI}}(\boldsymbol{r})\mathcal{M}_{xz}^{-1}
  \neq& H_{\mathrm{QAHI}}(\boldsymbol{r}), \\
\mathcal{T} H_{\mathrm{QAHI}}(\boldsymbol{r})\mathcal{T}^{-1}
  \neq& H_{\mathrm{QAHI}}(\boldsymbol{r}),
\end{align}
\begin{align}
  \mathcal{M}_{xz} =
    \left[
    \begin{array}{cc}
    i\hat{\sigma}_2 R_y    &
    0                \\
    0                &
    i\hat{\sigma}_2 R_y 
    \end{array}
    \right]  
  , \quad
  \mathcal{T} =
    \left[
    \begin{array}{cc}
    i\hat{\sigma}_2 \mathcal{K}  &
    0                \\
    0                &
    i\hat{\sigma}_2 \mathcal{K} 
    \end{array}
    \right], 
\end{align}
 where $R_y$ is the reflection operator about the $xz$ plane, (i.e. $y \rightarrow -y$) and $\mathcal{K}$ donates the complex 
 conjugation. However, the Hamiltonian preserves magnetic mirror reflection symmetry (MMRS) with respect to the $xz$-plane
 which is defined by combination of $M_{xz}$ and $\mathcal{T}$ as
\begin{align}
  \mathcal{T}_{xz} H_{\mathrm{QAHI}}(\boldsymbol{r})\mathcal{T}_{xz}^{-1}
  = H_{\mathrm{QAHI}}(\boldsymbol{r}), \label{mmmr_q}\\
  \mathcal{T}_{xz} =
    \left[
    \begin{array}{cc}
    \hat{\sigma}_0 R_y \mathcal{K} &
    0                \\
    0                &
    \hat{\sigma}_0 R_y \mathcal{K}
    \end{array}
    \right],
\end{align}
 where $\mathcal{T}_{xz}$ is the magnetic mirror reflection symmetry operator. 
 By applying $\mathcal{T}_{xz}$ to Eq.~(\ref{LS}), we find
\begin{align}
\mathcal{T}_{xz} H_{\mathrm{L}}(\boldsymbol{r})\mathcal{T}_{xz}^{-1}
  =& \left. H_{\mathrm{L}}(\boldsymbol{r}) \right|_{\theta_L \to -\theta_L}, \\
\mathcal{T}_{xz} H_{\mathrm{R}}(\boldsymbol{r})\mathcal{T}_{xz}^{-1}
  =& \left.H_{\mathrm{R}}(\boldsymbol{r})\right|_{\theta_R\to -\theta_R}.
\end{align}
As a consequence, we conclude that  
\begin{align}
\mathcal{T}_{xz} H_0(\theta,\boldsymbol{r})\mathcal{T}_{xz}^{-1}
  =& H_0(-\theta,\boldsymbol{r}). \label{sel}
\end{align}

The Bogoliubov-de Gennes equation can be described as, 
\begin{align}
H_0(\theta) \, \psi_n =& E_n(\theta) \, \psi_n, \label{BdG}
\end{align}
 where $\psi_n$ and $E_n$ are an eigenstate and an eigenvalue labeled by an index $n$, respectively. 
 By using Eq.~(\ref{sel}), the BdG equation can be transformed into 
\begin{align}
H_0(-\theta)\,  \mathcal{T}_{xz}\,  \psi_n =& E_n(\theta) \, \mathcal{T}_{xz} \, \psi_n.\label{BdGm}
\end{align}
From Eqs.~(\ref{BdG}) and (\ref{BdGm}), we conclude that $H_0(\theta)$ and $H_0(-\theta)$ 
have exactly the the same eigenvalues. Namely the relation
\begin{align}
E_n (\theta) =  E_n (-\theta).\label{En}
\end{align}
hold true in a S/QAHI/S junction.
Generally speaking, the energy of the Josephson junction $F(\theta)$ and the Josephson current are 
related to each other as
\begin{align}
F(\theta) =&
  \sum_n E_n f_F (E_n), \label{F} \\
J(\theta) =& \dfrac{2e}{\hbar} \dfrac{\partial F(\theta)}{\partial \theta},\label{jc}
\end{align} 
 where $f_F (E_n)$ is the Fermi distribution function. 
Due to Eq.~(\ref{En}), the energy of the junction is an even function of the phase difference $\theta$ 
and the Josephson current is an odd function of $\theta$. 
Therefore, the CPR satisfies
\begin{align}
J(\theta) = - J(- \theta),\label{jodd}
\end{align}
which indicates $J(\theta=0) = 0$. 
Thus, MMRS prohibits the appearance of the phase shift $\varphi$ in CPR in the absence of an external 
Zeeman field. 

The effects of the Zeeman field depends on its direction.
It is easy to confirm the follwing relations
\begin{align}
\mathcal{T}_{xz} H_{V_x} \mathcal{T}_{xz}^{-1} = H_{V_x}, \\
\mathcal{T}_{xz} H_{V_y} \mathcal{T}_{xz}^{-1} \neq H_{V_y}.
\end{align}
We find that the Zeeman potential $V_y$ breaks MMRS for the $xz$-plane, whereas $V_x$ preserves it.
As shown in Fig.~\ref{fig:main} (c), 
the phase shift is zero even in the presence of $V_x$.
On the other hand, the phase shift is proportional to $V_y$ as shown in 
Fig.~\ref{fig:main} (b).
The symmetry analysis of Hamiltonian explains well the anisotropic response of the phase shift 
to the direction of a Zeeman field. 
We conclude that breaking MMRS is a necessary condition for realizing a $\varphi$-junction.

The important point of the symmetry analysis can be understood in more phenomenological way. 
The relation between the two free energies $F(\theta)$ and $F(-\theta)$ determines the 
junction property. 
When $F(\theta)=F(-\theta)$ is satisfied,
one can immediately conclude that the junction is either 0- or $\pi$-junction. 
 In the two superconductors, the transformation of $\theta$ to $-\theta$ 
is realized by applying the complex conjugation to Eq.~(\ref{LS}).
Therefore, we find $F(\theta)=F(-\theta)$ if $\hat{Q}=\hat{Q}^\ast$ holds in Eq.~(\ref{Q}). 
The Hamiltonian in Eq.~(\ref{Q}) satisfies $\hat{Q}^\ast(x,y)=\hat{Q}(x,-y)$ as discussed in Eq.~(\ref{mmmr_q}).
Therefore, the junction may become $\varphi$-junction in the presence of the potential depending on the $y$ direction. 
We revisit this issue in Sec.~IV.
\subsection{Andreev reflection}

\begin{figure}[b]
  \begin{center}
  \includegraphics[clip, scale=0.4]{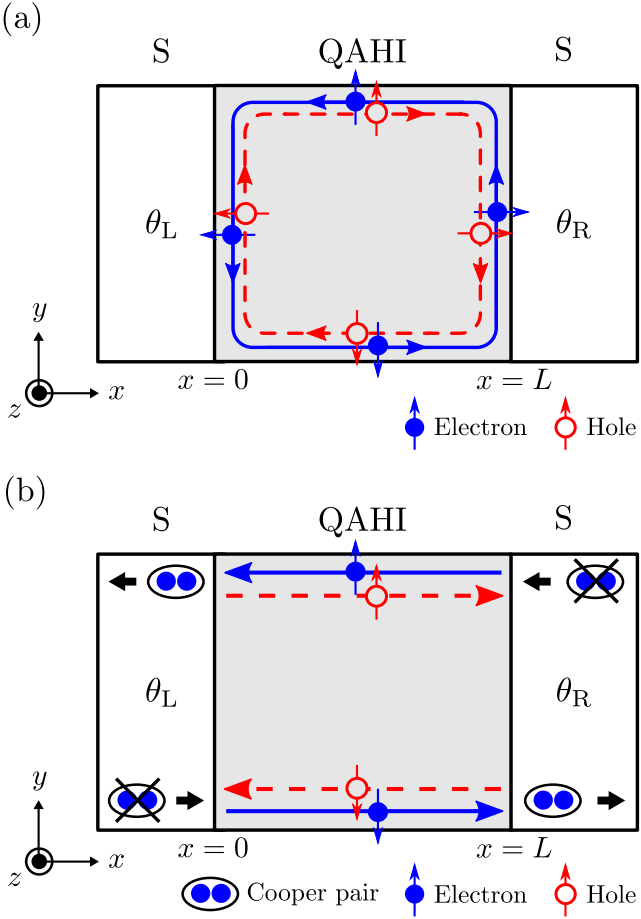}
  \end{center}
\caption{(a) The chiral edge states in an electron branch and a hole branch. The spin of a quasiparticle always
points outside a QAHI.
(b) The two Andreev reflection processes contribute to the lowest order Josephson coupling.
}\label{fig:arp}
\end{figure}
To explain the linear relation between $\varphi$ and $V_y$, we analyze the Andreev reflection processes 
which carry the Josephson current in a S/QAHI/S junction.
The chiral edge current flows along two surfaces and two interfaces to the $s$-wave superconductors as illustrated 
in Fig.~\ref{fig:arp}(a). An electron (a hole) moves in the counterclockwise (clockwise) direction.  
The direction of spin is locked to the direction of a quasiparticle motion and always points outwardly.
The spin-orbit interaction changes the spin direction of a quasiparticle at four corners of a QAHI.

When we focus on the edge states at the botom surface around $y=0$, the pair potential in the superconductors hybridizes 
an electron and a hole near the junction interfaces at $x=0$ and $L$,
which causes the Andreev reflection as shown in Fig.~\ref{fig:arp}(b). 
The amplitude of the Andreev reflection, however, is expected to be very small due to the spin mismach 
in the reflection processes. 
Usually, a spin-singlet superconductor causes the Andreev reflection which converts a spin $\uparrow$ ($\downarrow$) 
electron into a spin $\downarrow$ ($\uparrow$) hole. 
However, at the bottom edge, spin $\uparrow$ channels are absent in both electron and hole branches.
In the edge states at the top surface around $y=W$, the spin $\downarrow$ channels are absent.
The spin mismach drastically suppresses the Josephson current.
Actually the amplitude of the Josephson current in Fig.~\ref{fig:main} is much smaller than 
$J_0$ even in the absence of potential barrier at the interface. 
Although we have tried to analyze the spin-flip Andreev reflection process in the presence of spin-orbit 
coupling, we cannot derive a simple analytic expression of the Andreev reflection coefficients. 
\begin{figure}[t]
  \begin{center}
  \includegraphics[clip, scale=0.42]{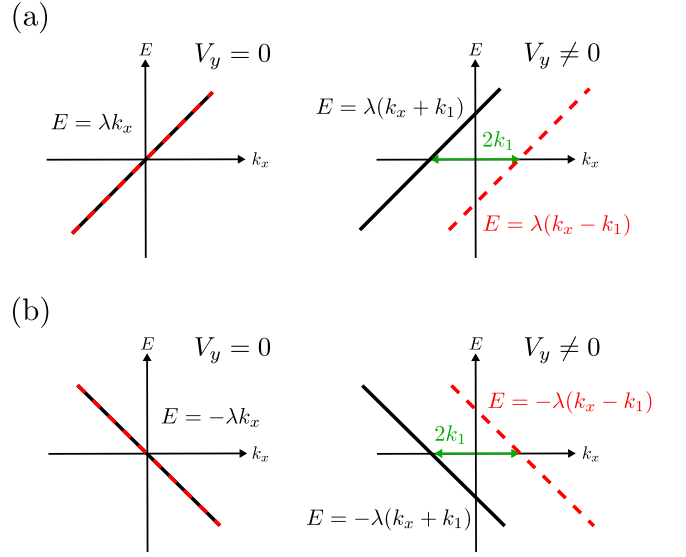}
  \end{center}
\caption{The dispersion relation of edge state at the bottom (a) and that at the top (b). 
Solid and broken lines are the dispersion in an electron branch and a hole branche, respectively.}\label{fig:dsp}
\end{figure}
%
The numerical results, however, suggest the presence of spin-flip Andreev reflection at the interfaces.
Therefore we assume the spin-flip Andreev reflection at the two interfaces and describe 
the phase shift of a quasiparticle in the reflection processes.

Figure~\ref{fig:arp}(b) shows two Andreev reflection processes which contribute to the lowest order Josephson coupling.
In the botom edge, an electron moves to the right and a hole moves to the left, which carry a Cooper pair 
from the left superconductor to the right superconductor. 
At the top edge, a hole moves to the right and an electron moves to the left. Such process carry a Cooper pair 
from the right superconductor to the left.
The Josephson current is described by the subtraction of the two reflection processes. 
A current formula~\cite{Y.Asano} may describe the Josephson current in such situation phenomenologically,
\begin{align}
J =& \frac{ie}{\hbar}
   T     
   \sum_{\omega_n} 
   [
   {r}^{eh}_L\,  \cdot \, {t}^h_B \, \cdot\, {r}^{he}_R \,  \cdot {t}^e_B \nonumber \\
   & -\hat{r}^{he}_L\, \cdot\, {t}^e_T\, \cdot\,  {r}^{eh}_R\, \cdot \, {t}^h_T
     ], \label{jcph}
\end{align}
where $r^{he}_{L(R)}$ ($r^{eh}_{L(R)}$) is the spin-flip Andreev reflection coefficient 
at the left (right) interface from the electron branch to the hole branch
(from the hole branch to the electron branch) and $t^e_{B(T)}$ ($t^h_{B(T)}$) is the 
transmission coefficient of an electron ( a hole) along the bottom (top) edge.
The first (second) term corresponds to the Andreev reflection process at the bottom (top) edge.
The Andreev reflection coefficients are phenomenologically described by 
\begin{align}
r^{he}_{L(R)} =& -i \frac{\Delta}{\Omega} \sqrt{t_I} e^{-i\theta_{L(R)}},\\
r^{eh}_{L(R)} =& -i \frac{\Delta}{\Omega} \sqrt{t_I} e^{i\theta_{L(R)}},
\end{align} 
where $t_I$ is the effective transmission probability with spin-flipping at the interface,
 and $\Omega=\sqrt{\omega_n^2+\Delta^2}$.

The dispersion relation of the edge state at the bottom and that at the top are shown in Fig.~\ref{fig:dsp} (a) and (b), 
respectively. The results are analytically obtained by solving the BdG Hamiltonian near the edges. See also Appendix A for details.
In the absence of $V_y$, the dispersion in the electron branch and that in the hole branch are identical to each other. 
The wave number in the $x$ direction at the Fermi level is zero. 
Thus the transmission coefficients would be described by
$t^{e}_B=t^h_B=t^e_T=t^h_T=t_0$ with $t_0$ being a real number.
The resulting Josephson current in Eq.~(\ref{jcph}) at $T=0$ becomes
\begin{align}
J= \frac{e \Delta}{\hbar} \, t_0^2\,  t_I \, \sin \theta.
\end{align}

When we introduce the Zeeman potential in the $y$ direction, 
the electron dispersion becomes $E=\lambda(k_x+k_1)$, whereas 
the hole dispersion becomes $E=\lambda(k_x-k_1)$ at the botom edge as shown in Fig.~\ref{fig:dsp} (a). 
As shown in Fig.~\ref{fig:dsp} (b), the electron and hole dispersion at the top edge 
are deformed as $E=-\lambda(k_x+k_1)$ and $E=-\lambda(k_x-k_1)$, respectively. 
As a consequence, the transmission coefficients through the edge state 
should be modified as $t^e_B = t_0 e^{-ik_1L}$, $t^h_B = t_0 e^{-ik_1 L}$, $t^e_T = t_0 e^{ik_1L}$, $t^h_T = t_0 e^{ik_1L}$.
The Josephson current in Eq.~(\ref{jcph}) in such situation becomes 
\begin{align}
J=& \frac{e \Delta}{\hbar}\, t_0^2\,  t_I\,  \sin(\theta - \varphi),\\
\varphi=& 2k_1 L = \frac{2V_y L}{\lambda}. \label{shift_v}
\end{align}
The phenomenological argument explains the linear relation between the phase shift $\varphi$ on the Zeeman potential $V_y$. 
In Fig.~\ref{fig:main} (b), we plot $\varphi$ in Eq.~(\ref{shift_v}) with a solid line. 
The phenomenological results in Eq.~(\ref{shift_v}) explain the numerical results even quantitatively. 
A Zeeman field in the $y$ direction affects mainly the wave numbers at the edge states. 
As a result, the amplitude of the Josephson current is independent of $V_y$ as shown in Eq.~(\ref{shift_v}) and 
in numerical simulation in Fig.~\ref{fig:main} (a). 
The perfect agrement between Eq.~(\ref{shift_v}) and the numerical results suggests the validity 
of the phenomenological argument.

\section{Another $\varphi$-junctions}

The symmetry analysis in Sec.~III suggests that the breakdown of MMRS is a trigger of a $\varphi$-junction. 
To check the validity of the conclusion, 
we study the effects of breaking MMRS by 
another physical sources such as (i) random impurity potential in a QAHI 
 and (ii) asymmetric junction geometry with respect to the $xz$-plane. 
In this section, we set the Zeeman potentials to be zero as $V_x =V_y = 0$. 

\subsection{Impurity potential} 
\begin{figure}[H]
  \begin{center}
    \includegraphics[clip,scale=0.75]{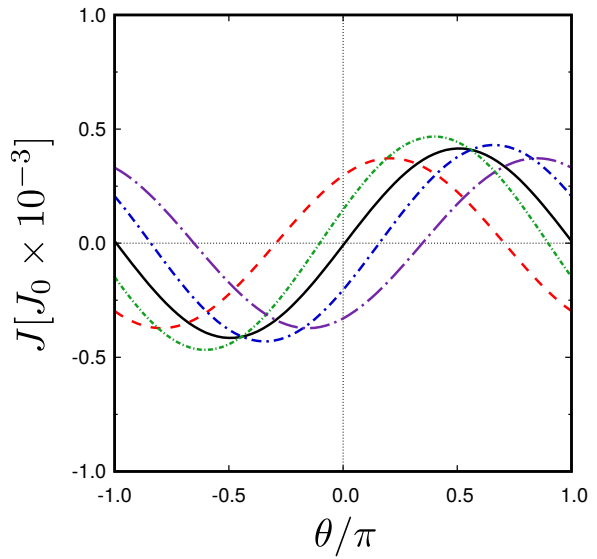}
    \caption{The Josephson currents for four samples with different impurity configurations are plotted as a function of 
$\theta$ with broken lines. The ensemble average of the Josephson current over 1500 samples is shown with a solid line.}
    \label{fig:cpr-imp}
  \end{center}
\end{figure}

The impurity potentials in a QAHI ($1 \leq j \leq L$ ) are considered through the random on-site potential 
$V_{\mathrm{imp}}(\boldsymbol{r})\, \delta_{\boldsymbol{r},\boldsymbol{r'}}$, 
where $V_{\mathrm{imp}}(\boldsymbol{r})$ is given randomly in the range of $-V_D/2 \leq V_{\mathrm{imp}}(\boldsymbol{r}) \leq V_D/2$. 
Equation~(\ref{QR}) is transformed as $ \hat{Q}(\boldsymbol{r}) \rightarrow \hat{Q}(\boldsymbol{r}) 
+ V_{\mathrm{imp}}(\boldsymbol{r}) \hat{\sigma}_{0}$. 
The impurity potential breaks MMRS because
\begin{align}
R_y V_{\mathrm{imp}}(x,y) R_y^{-1}=V_{\mathrm{imp}}(x,-y) \neq V_{\mathrm{imp}}(x,y),  
\end{align}
due to its random character.
 In Fig.~\ref{fig:cpr-imp}, we show the CPR of the Josephson current in the presence of impurity potential 
with $V_D = 0.25t$. The four CPR's with broken lines correspond to the results 
for four samples with different random impurity configurations. 
 The Josephson current flows at $\theta=0$ in all the samples. 
Although the amplitude of the current is insensitive to the impurity configuration, the phase shift $\varphi_i$ 
for the $i$ th sample depends seriously on the random potential configuration. 
 We also plot the ensemble average of the Josephson current over 1500 samples with a solid line. 
The results show that the Josephson current after averaging recovers the sinusoidal CPR.
In experiments, however, the phase shift $\varphi$ is expected in a measurement of the Josephson current 
in a single sample because the Josephson effect is a phase coherent phenomenon. 
Thus the ensemble averaged Josephson current cannot predict a result of one-shot measurement in a single 
sample~\cite{Y.Asano_PRB_2001}.

\subsection{Asymmetric junction geometry} 
\begin{figure}[H]
  \begin{center}
    \includegraphics[clip,scale=0.50,]{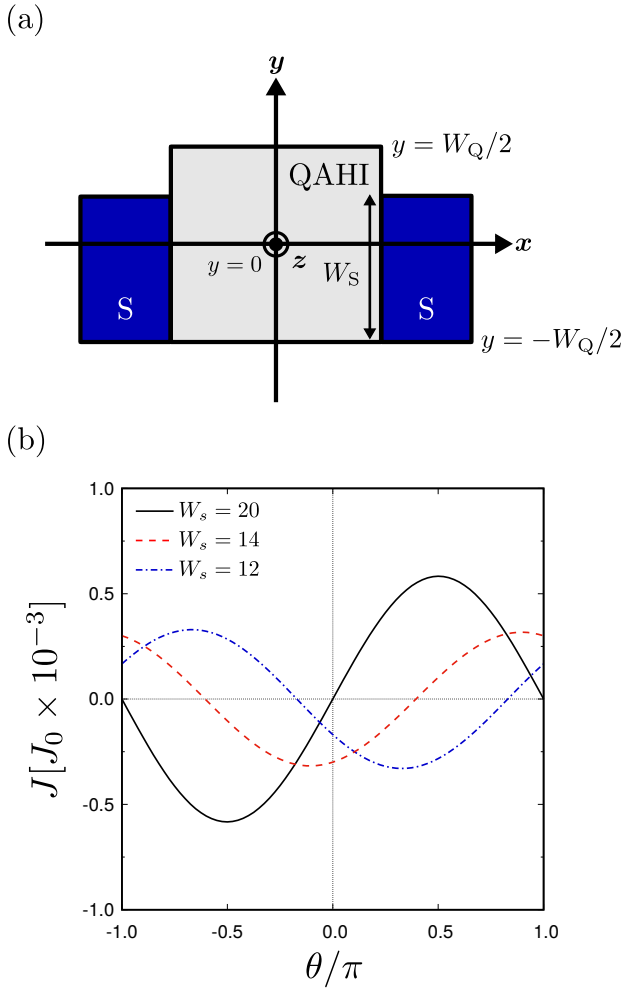}
    \caption{(a) A Josephson junction asymmetric with respect to the $xz$ plane. 
	The width of the two superconductors $W_{\mathrm{S}}$ is different from that of the QAHI $W_{\mathrm{Q}}$. 
   (b) The Josephson current is plotted as a function of $\theta$ for various $W_{\mathrm{S}}$.}
    \label{fig:asm}
  \end{center}
\end{figure}

A junction shown in Fig.~\ref{fig:asm} (a) is
asymmetric with regard to the $xz$-plane at $y=0$, where we choose the width of two superconductors as $W_{\mathrm{S}}$ and that 
of a QAHI as $W_{\mathrm{Q}}$. 
In such situation, the junction geometry breaks MMRS because the total Hamiltonian is no longer symmetric under $y \leftrightarrow -y$.
%
%
In Fig.~\ref{fig:asm} (b), the CPR of the Josephson effect for various $W_{\mathrm{S}}$ is represented at $W_{\mathrm{Q}} = 20$. 
The results for $W_{\mathrm{S}}=14$ and 12 show the phase shift in the CPR by breaking down of MMRS due to the two superconducting lead wires. 
At $W_{\mathrm{S}}=W_{\mathrm{Q}}$, the phase shift becomes zero as shown with a solid line. 
Within our numerical simulation, however, we cannot find any systematic relation between $W_{\mathrm{S}}/W_{\mathrm{Q}}$ and 
the phase shift $\varphi$. 
 
In experiments, it is almost impossible to controle both the junction geometry and the impurity configuration. 
Therefore the phase shift in the CPR always can be expected in every sample. 
This feature is peculiar to the Josephson junction consisting of a QAHI.

\section{Conclusion}
We have studied the direct-current Josephson effect through a quantum anomalous Hall insulator (QAHI)
 by using the recursive Green function method. 
A QAHI breaks both time-reversal symmetry and mirror reflection symmetry simultaneously. 
However, their combined symmetry called magnetic mirror reflection symmetry is preserved. 
The current-phase ($J-\theta$) relationship (CPR) in the Josephson effect shows very unusual features by reflecting 
such symmetry property of a QAHI. 
In the presence of magnetic mirror reflection symmetry, the CPR is always sinusoidal as usual, (i.e., $J\propto \sin\theta$). 
In-plane Zeeman fields, random impurities in the QAHI, and asymmetric junction geometries 
break magnetic mirror reflection symmetry. As a consequence, the CPR deviated from the sinusoidal 
relation to $J\propto(\theta-\varphi)$. 
The phase shift $\varphi$ by impurities and that by the asymmetric junction geometry would be 
out of contorole in experiments.
On the other hand, the phase shift is proportional to an in-plane Zeeman field. 
By considering the Andreev reflection processes phenomenologically, we explain the linear 
relationship between the phase shift and a Zeeman field. 
We conclude that $\varphi$ is tunable in a Josephson junction consisting of a QAHI.

%
%
%
\begin{acknowledgments}
We are grateful to T.~Habe, S.-I.~Suzuki, Ya.~V.~Fominov and A.~A.~Golubov for helpful discussion. 
This work was supported by Topological Materials Science (No.~JP15H05852) and 
KAKENHI (No.~JP15H03525) from the Ministry of Education, Culture, Sports, Science and Technology (MEXT) of 
Japan, JSPS Core-to-Core Program (A. Advanced Research Networks), Japan-RFBR JSPS Bilateral Joint
Research Projects/Seminars (2717G8334b),
and by the Ministry of Education and Science of the Russian Federation
(Grant No.~14Y.26.31.0007). 
S. I. is supported in part by a Grant-in-Aid
for JSPS Fellows (Grant No. JP16J00956) provided by the
Japan Society for the Promotion of Science (JSPS).
\end{acknowledgments}

\appendix


\section{Chiral edge states} 

 We represent the analytic expression of wave function and energy dispersion 
of chiral edge states at a surface of a QAHI in the presence of a Zeeman field. 
 Eq.~(\ref{QAHI}) in momentum space is given by
\begin{align}
 H_{\mathrm{QAHI}}
    =&
    \left[
    \begin{array}{cc}
    \hat{Q}(\boldsymbol{k})  &
    0                \\
    0                &
    -\hat{Q}^* (-\boldsymbol{k})  
    \end{array}
    \right], \label{QAHIff} \\
  \hat{Q} (\boldsymbol{k})
    =& \varepsilon_{\boldsymbol{k}} \hat{\sigma}_3
    -\lambda k_x \hat{\sigma}_2
    +\lambda k_y \hat{\sigma}_1 - V_y \hat{\sigma}_2,\\
  \varepsilon_{\boldsymbol{k}} =& \dfrac{\hslash^2 }{2m}
       {(\boldsymbol{k}^2-k_0^2)},
\end{align}
where $k_0=\sqrt{2mm_z/\hbar^2}$ is derived from
the spontaneous magnetization in the $z$ direction.
Since the pair potential is absent, the BdG equation is decoupled into two equations,
\begin{align}
    \hat{Q}(\boldsymbol{k})\psi^e =&
    E \psi^e, \\
    -\hat{Q}^*(-\boldsymbol{k})\psi^h =&
    E \psi^h.    
\end{align}
In the electron branch, the energy dispersion and the wave functions are obtained as
\begin{align}
      \left(
        \begin{array}{c}
        E_{\boldsymbol{k}}^e + \varepsilon_{\boldsymbol{k}} \\
        \lambda(k_y - i\tilde{k}_x) 
        \end{array}
      \right), \quad 
      \left(
        \begin{array}{c}
        \lambda(k_y + i\tilde{k}_x) \\
        E_{\boldsymbol{k}}^e - \varepsilon_{\boldsymbol{k}} 
        \end{array}
      \right)
\end{align}
for $E_{\boldsymbol{k}}^e$ and $-E_{\boldsymbol{k}}^e$, respectively. Here we define following quantities,
\begin{align}
E_{\boldsymbol{k}}^e =&
    \sqrt{\varepsilon_{\boldsymbol{k}}^2 + \lambda^2 
\left\lbrace  \left( k_x + k_1 \right)^2 + k_y^2)\right\rbrace },  \\
\boldsymbol{k} =& k_x +k_1, \quad
 k_1 = V_y / \lambda.
\end{align}
At $V_y = 0$, the spin-orbit interaction and the magnetic moment induce an energy gap $E_{\mathrm{g}} = 2\lambda k_0$. 
In a weak external Zeeman field $V_y \ll m_z$, the energy gap remains finite and chiral edge states appear.

 In order to obtain the wave function and the dispersion of the chiral edge state, 
 we consider a semi-infinite system that has a surface perpendicular to the $y$-axis at $y=0$.
 At an energy $E>0$, the wave function of the edge states $  \psi_{\mathrm{Q}}^e(y) {e}^{ik_x x}$ is calculated to be
\begin{align}
   \psi_{\mathrm{Q}}^e(y) =& 
      \left(
        \begin{array}{c}
        f_{+} \\
        \gamma_{k_{+},-} 
        \end{array}
      \right)
      \mathrm{e}^{ik_+ y} A_+ + 
     \left(
       \begin{array}{c}
       \gamma_{-k_-,+}  \\
       f_{-}
       \end{array}
     \right)
     \mathrm{e}^{-ik_- y}A_-,
\end{align}
\begin{align}
  f_{\pm} = &
    E + \sqrt{D^e} \mp \lambda k_0 \tilde{\lambda}/2, \quad
  \gamma_{k,\pm} = \lambda(k \pm ik_x), \\
  k_{\pm} =& \sqrt{A_0 \pm (2m/\hbar^2)\sqrt{D^e}}, \\
  D^e =& E^2 - \lambda^2 
       \left\lbrace(k_0^2-k_x^2 
       + \left( k_x + k_1 \right)^2 \right\rbrace  \\ \nonumber
     & + (\tilde{\lambda} k_0)^2 \lambda^2/4, \\
   \tilde{\lambda} =& \lambda k_0 / \varepsilon_0, \quad
   \varepsilon_0 = \hbar^2 k_0^2 /2m, \\
   A_0 =& k_0^2 -k_x^2 -(\tilde{\lambda}k_0)^2 /2,
 \end{align}
where $A_{\pm}$ are the amplitudes of the wave function. 
The condition $D^e<0$ results in the complex wave number in the $y$ direction.
By imposing the boundary condition $\psi_{Q}^e(0)=0$, we obtain the dispersion of chiral edge state as 
\begin{align}
E_{BS}^e=\lambda (k_x + k_1),
\end{align}
for $k_x^2<k_0^2(1-\tilde{\lambda}^2/4)$ and the wave function as
\begin{align}
   \psi_{\mathrm{ES}}^e(y) =& 
      C_0 \mathrm{e}^{-y/\xi}
      \sin{y\sqrt{k_0^2 -k_x^2}} 
      \left(
        \begin{array}{c}
        \delta_0 \\
        \delta_0^* 
        \end{array}
      \right), \label{EES} \\
   \xi = & \dfrac{\hbar^2}{m\lambda},
 \end{align}
where $\delta_0 = \mathrm{e}^{i \pi/4}$ and $\xi$ is a localization length of the edge states. 
We have used a relation $\tilde{\lambda}\ll 1$. 
 The dispersion and the wave function in the hole brach can be obtained in a similar way as
\begin{align}
E_{BS}^h =& \lambda (k_x - k_1), \\
   \psi_{\mathrm{ES}}^h(y) =& 
      C_0 \mathrm{e}^{-y/\xi}
      \sin{y\sqrt{k_0^2 -k_x^2}} 
      \left(
        \begin{array}{c}
        \delta_0^* \\
        \delta_0 
        \end{array}
      \right), \label{HES}      
\end{align}
 for $k_x^2<k_0(1-\tilde{\lambda}\lambda/4)$.

The current of the probability density in the $x$ direction are represented by
\begin{align}
  J_x^e = \dfrac{\lambda}{\hbar}
          C_0^2 \sin^2y\sqrt{k_0^2-k_x^2}
          \exp^{-2y/\xi} ,\label{JXE} \\
  J_x^h = -\dfrac{\lambda}{\hbar}
          C_0^2 \sin^2y\sqrt{k_0^2-k_x^2}
          \exp^{-2y/\xi},\label{JXH} 
\end{align}
for an electron and a hole, respectively. 
Thus, an electron (a hole) moves to the right (left) direction at the bottom edge of a QAHI as shown in Fig.~\ref{fig:arp}.

\section{Josephson current in the absence of Zeeman field}

We display numerical results of Josephson current in the absence of Zeeman field ($V_x=V_y= 0.0$), to discuss 
unusual Josephson effect in a S/QAHI/S junction. 
As we explained in the text, several parameters are fixed also in the Appendix as $W=20$, $L=80$,  
 $m_z=1.5t$, $\lambda=1.0t$, and $\Delta_0=0.001t$.

In Fig.~$7$, we plot the Josephson current as a function of $\theta$ for several choices of $\mu_s$. 
In all cases, the CPR at $T=0.1 T_c$ is sinusoidal in the absence of the Zeeman field.
\begin{figure}[h]
  \begin{center}
    \includegraphics[clip,scale=0.7]{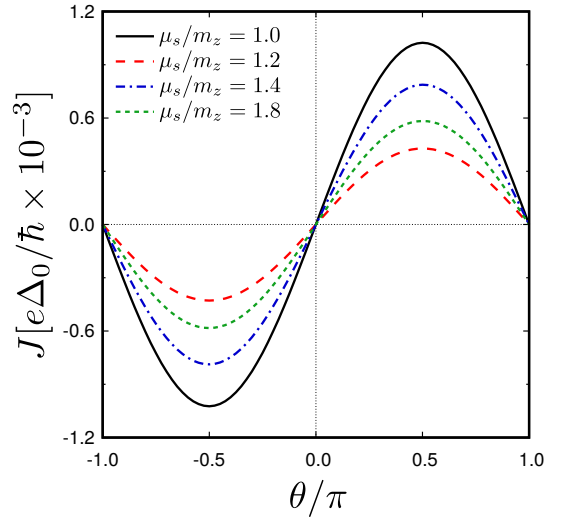}
\caption{The Josephson current versus $\theta$ for several choices of $\mu_s$ at $T=0.1T_c$. 
}
\end{center}
\label{mus_dep}
\end{figure}

Unseal features can be seen in the dependence of the Josephson critical current on temperatures as shown in 
Fig.~$8$. In all cases, the CPR is sinusoidal as discussed in Fig.~$7$. 
The critical current shows the nonmonotonic dependence on temperatures and takes its maximum around $T\approx 0.08 T_c=\epsilon_0$.
The results suggest the existence of resonant-like state at $\epsilon_0$. Unfortunately, however, we cannot figure out physical reasons 
of such subgap states at the edges of a QAHI. The results for $\mu_s/m_z=1.4$ and tose for 1.8 have minimum around $T=0.2T_c$.
At present, the reasons of such unusual behavior are open question.
\begin{figure}[H]
  \begin{center}
    \includegraphics[clip,scale=0.7]{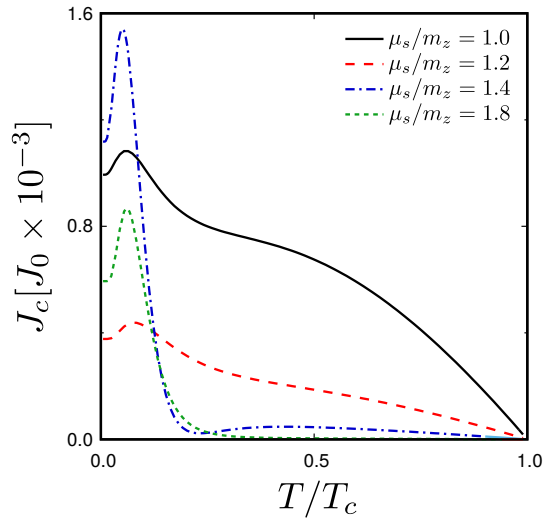}
\caption{The critical current $J_c$ versus temperature $T$ for various chemical potentials.
}
\end{center}
\label{tdep-1}
\end{figure}

Figure~$9$ shows the critical current versus $\mu_s$. 
The results for $T=0.1T_c$ correspond to the resonant-like peak in Fig.~$8$.
At a low temperature $T=0.1T_c$, the results show 
an aperiodic oscillating behavior as a function of $\mu_s$ and becomes almost zero around $\mu_s/m_z=1.1$, 1.6 and 1.9. 
The results also suggests existence of a resonant-like subgap states at the edge of a QAHI.

\begin{figure}[H]
  \begin{center}
    \includegraphics[clip,scale=0.7]{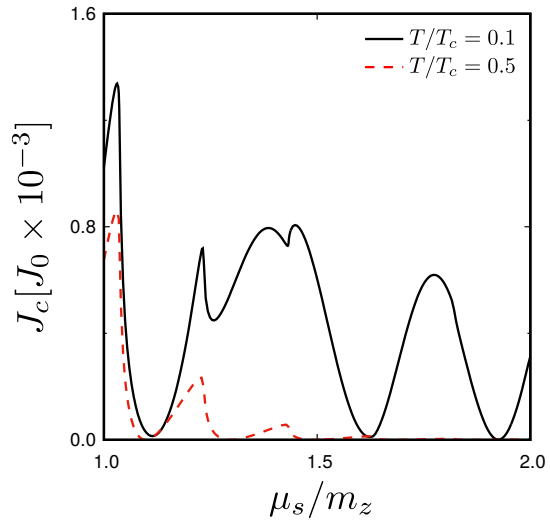}
\caption{The critical current versus $\mu_s$ for $T = 0.1T_c$ and $T = 0.5T_c$.
}
\end{center}
\label{mudep-jc}
\end{figure}

\end{document}